\documentclass[12pt]{article}
\usepackage[colorlinks,linkcolor=blue,citecolor=blue,urlcolor=blue,bookmarks,bookmarksnumbered]{hyperref}
\usepackage[paper=letterpaper,margin=1.0in]{geometry}
\usepackage{graphicx}
\usepackage{xcolor}

\usepackage{latexsym}
\usepackage{revsymb}
\usepackage{amsmath,amssymb,amsfonts,amsthm}
\usepackage{bm}
\usepackage{subfigure}
\usepackage{cite}
\usepackage{color}

\newcommand{\be}{\begin{equation}}
\newcommand{\ee}{\end{equation}}
\newcommand{\bea}{\begin{eqnarray}}
\newcommand{\eea}{\end{eqnarray}}

\newcommand{\X}{\mathbb{X}}
\newcommand{\hX}{\hat{\mathbb{X}}}
\newcommand{\ha}{\mathord{\hat a}}
\newcommand{\hx}{\mathord{\hat x}}
\newcommand{\tx}{\mathord{\tilde x}}
\newcommand{\htx}{\mathord{\hat{\tilde x}}}
\newcommand{\TeV}{\text{T\kern0pte\kern-1ptV}}
\newcommand{\Edual}{\tilde{E}}

\renewcommand{\thefootnote}{\fnsymbol{footnote}}

\begin{document}

${}$

%\title{Infinite Statistics, Dark Energy and the $H_0$ Tension}
%\author{Vishnu Jejjala\textsuperscript1} 
%\author{Michael Kavic\textsuperscript2} 
%\author{Djordje Minic\textsuperscript3} 
%\author{Tatsu Takeuchi\textsuperscript3} 
%\affiliation{\textsuperscript1Department of Physics and Astronomy, Wits, South Africa.} 
%\affiliation{\textsuperscript2Department of Physics and Astronomy, SUNY,  U.S.A.} 
%\affiliation{\textsuperscript3Department of Physics, Virginia Tech, Blacksburg, VA 24061, U.S.A.} 

\noindent
\begin{center}
\vspace{7mm}
{\Large\bfseries Dynamical Dark Energy and Infinite Statistics}
\vspace{7mm}

\textbf{
	Vishnu Jejjala${}^{1}$\footnote{\href{mailto:vishnu@neo.phys.wits.ac.za}{\texttt{vishnu@neo.phys.wits.ac.za}}},
	Michael J. Kavic${}^{2}$\footnote{\href{mailto:kavicm@oldwestbury.edu}{\texttt{kavicm@oldwestbury.edu}}},
    Djordje Minic${}^{3}$\footnote{\href{mailto:dminc@vt.edu}{\texttt{dminic@vt.edu}}},
and
	Tatsu Takeuchi${}^{3}$\footnote{\href{mailto:takeuchi@vt.edu}{\texttt{takeuchi@vt.edu}}}
}
\vspace{3mm}

{\footnotesize\it
${}^1$Mandelstam Institute for Theoretical Physics, School of Physics, NITheP, and CoE-MaSS,\\
University of the Witwatersrand, Johannesburg, WITS 2050, South Africa\\
${}^2$Department of Chemistry and Physics, SUNY Old Westbury, Old Westbury, NY 11568, U.S.A.\\
${}^3$Department  of Physics, Virginia Tech, Blacksburg, VA 24061, U.S.A.\\
}
\end{center}

%\date{\today}

\vspace{.5cm}

\begin{abstract}%\unskip
\noindent
In the $\Lambda$CDM model, dark energy is viewed as a constant vacuum energy density, the cosmological constant in the Einstein--Hilbert action.
This assumption can be relaxed in various models that introduce a dynamical dark energy.
In this letter, we argue that the mixing between infrared and ultraviolet degrees of freedom in quantum gravity lead to infinite statistics, the unique statistics consistent with Lorentz invariance in the presence of non-locality, and yield a fine structure for dark energy.
Introducing IR and UV cutoffs into the quantum gravity action, we deduce the form of $\Lambda$ as a function of redshift and translate this to the behavior of the Hubble parameter.

\end{abstract}

%\pacs{}

%\maketitle

\newpage

\renewcommand{\thefootnote}{\arabic{footnote}}
%%%%%%%%%%%%%%%%%%%%%%%%%%%%%%%%%%%%%%%%%%%%%%%%%%%%%%%%%%%%%%%%%%%%%%%%%%%%%% 
\section{Introduction} 

Since the seminal discovery that the expansion of our Universe is currently accelerating \cite{Riess:1998cb,Perlmutter:1998np},
many models have been proposed to explain and understand this observation \cite{Peebles:2002gy,Copeland:2006wr}.
These include various dark energy models in which
the acceleration is due to a currently small but non-zero vacuum energy density, quintessence models which incorporate inflaton-like scalar fields, and
$f(R)$ modified gravity models in which late-time acceleration is
due to subdominant terms in the action that become important at small curvature \cite{Nojiri:2006ri}.
See also \cite{Dent:2011zz}.

In the widely accepted $\Lambda$CDM cosmology, dark energy is provided by the time independent cosmological constant $\Lambda$ \cite{Weinberg:1988cp}. 
However, recent disagreements between the values of the Hubble constant $H_0$ 
determined from early ($z\sim 1000$ \cite{Aghanim:2018eyx}) and late ($z< 10$ \cite{Verde:2019ivm,Riess:2020sih,Pesce:2020xfe,deJaeger:2020zpb,Schombert:2020pxm}) Universe observations
have rekindled interest in dynamical dark energy models in which
the vacuum energy density is time dependent
\cite{quintessence,Alam:2004jy,Joyce:2014kja,Zhao:2017cud,Sola:2016ecz,Sola:2018sjf,Yang:2018qmz,Poulin:2018cxd,Pan:2019gop}.

Furthermore, it has also been argued that the quantization of gravity may naturally lead to
time dependent dark energy \cite{Polyakov:1993tp,Kitamoto:2019rij,Berglund:2019ctg}.
In particular, Ref.~\cite{Berglund:2019ctg} points out that the 
matrix model formulation of non-perturbative quantum gravity 
proposed in Refs.~\cite{Freidel:2013zga,Freidel:2015uug,Freidel:2016pls,Freidel:2017xsi,Freidel:2017wst}
naturally leads to a time dependent $\Lambda(z)$
due to the doubling of spacetime that is required in this approach.
There, the cosmological constant in the observable spacetime is given by the integration over the 
unobservable dual spacetime curvature, and consequently, the dynamical evolution of the latter leads to the
time dependence of the former.

In this letter, we follow the line of thought of Ref.~\cite{Berglund:2019ctg} and propose a concrete 
functional form for $\Lambda(z)$ by focusing on another feature of quantum gravity, 
namely infinite statistics \cite{Greenberg:1989ty,Strominger:1993si,Jejjala:2006jf,Jejjala:2007hh,Jejjala:2007rn}.
While we work within one particular framework for definiteness, we emphasize that much of what we say is a generic feature of theories in which UV and IR degrees of freedom mix.
This is natural in low energy effective field theories obtained from string theory, for example as a consequence of non-commutativity.

Infinite statistics is motivated by non-locality and Lorentz covariance,
and is realized in large-$N$ matrix models \cite{Gopakumar:1994iq}.
It is the statistics of the partonic degrees of freedom of the matrix theory from which spacetime is constructed,
and suggests a form for the density of states for the partons from
which $\Lambda(z)$ can be calculated. 
In other words, the functional form of $\Lambda(z)$ we propose is a manifestation of the
infinite statistics that the partons must follow.
This is similar to other macroscopic statistical manifestations of quantum mechanics 
such as blackbody radiation.

%\red{(The main logic of this paper is as follows. Given that the new proposal for non-perturbative quantum gravity is a matrix model, in which, in a certain limit, dark energy is realized as a dynamical geometry of dual spacetime, and given the fact that matrix models, in the limit of the large matrix size, can be directly related to infinite statistics, we argue that infinite statistics can be used to model the fine structure of dark energy. In some sense, the partonic degrees of freedom of the quantum gravitational matrix model, out of which both spacetime and matter degrees of freedom are conjectured to emerge, obey infinite statistics, and thus, infinite statistics controls the structure of the vacuum energy/dark energy sector.)}

This letter is organized as follows.
In Section~2, we briefly review the new approach to quantum gravity based on 
``quantum relativity''~\cite{Freidel:2013zga, Freidel:2015uug, Freidel:2016pls, Freidel:2017xsi, Freidel:2017wst} 
and there we outline the general argument for the variation of $\Lambda(z)$ with time.
In Section~3, we review the relevance of infinite statistics to quantum gravity.
%
%and finally, we present an example for such a variation based on
%the recent proposal \cite{Berglund:2019ctg} that is rooted in a
%particular form of quantum statistics, called infinite statistics \cite{Greenberg:1989ty} 
%and macroscopic effects of quantum gravity 
%(see also, \cite{Strominger:1993si}).
Infinite statistics is realized in large-$N$ matrix models \cite{Gopakumar:1994iq}
and thus it is particularly appropriate for the new matrix model like approach reviewed in section~2.
%to non-perturbative quantum gravity based on 
%``quantum relativity''~\cite{Freidel:2013zga, Freidel:2015uug, Freidel:2016pls, Freidel:2017xsi, Freidel:2017wst}.
%Finally, %we apply these insights to the problem of the $H_0$ tension.
%We start by discussing the relevance of infinite statistics in quantum gravity, and 
%then we discuss the relevance of infinite statistics for the fine structure
%of dark energy. This in turn implies that the currently measured values
%of $H_0$ are all correct and they are indicative of such a fine structure in the dark energy sector. 
%In particular, 
In Section~4,  
we derive an explicit formula for $\Lambda(z)$, and the resulting
modification to the evolution of the Hubble parameter $H(z)$ is discussed
in Section~5,
This, in principle, can be compared with actual observation of $H(z)$.
Section~6 concludes with a few remarks.
%and discuss possible predictions
%which should be useful in understanding the present and
 %for the 
%upcoming observational searches.
%the value of 
%$H_0$ for the dark ages, and indicate that this value will be soon probed
%by the James Webb Space Telescope (JWST).
%Finally, we discuss the fundamental theoretical formulation of our proposal
%and end the letter with a few concluding comments.

%%%%%%%%%%%%%%%%%%%%%%%%%%%%%%%%%%%%%%%%%%%%%%%%%%%%%%%%%%%%%%%%%%%%%%%%%%%%%% 
\section{Quantum Spacetime and Quantum Gravity}

%In this section we remark on the fundamental realization of dark energy in quantum gravity and the underlying theoretical structure behind our discussion.
%The idea is that quantum gravity can be viewed as a gravitization of a geometry of quantum non-locality that is consistent with causality, which underlies any quantum theory (called Born geometry - the geometry of modular spacetime) \cite{Freidel:2013zga, Freidel:2015uug, Freidel:2016pls, Freidel:2017xsi, Freidel:2017wst}. 
%In that context, quantum spacetime is doubled and non-commutaive and dark energy in our observed classical spacetime can be viewed as a curvature of a dual spacetime \cite{Berglund:2019ctg}. 
%(Dual spacetime is conjugate, in the sense of canonical commutation relations, to our observed spacetime).
%Note that the crucial prediction of this approach to quantum gravity, which is fundamentally non-local and covariant \cite{Freidel:2013zga, Freidel:2015uug, Freidel:2016pls, Freidel:2017xsi, Freidel:2017wst}, is the occurrence of new physical phenomena in the deep infrared.
%Within this context dark matter is realized in terms of degrees of freedom dual to the observed matter degrees of freedom and dark energy is realized via the dynamical geometry of the dual spacetime, as explained in what follows \cite{Freidel:2017xsi}.
%In essence, here we are tracing the story of relativity from special to general, albeit in the quantum context, quantum theory (including quantum field theory) being understood as special quantum relativity and quantum  gravity being realized as general quantum relativity.

In this section, we outline the non-perturbative formulation of
quantum gravity in terms of a doubled matrix model quantum theory proposed in Refs.~\cite{Freidel:2013zga, Freidel:2015uug, Freidel:2016pls, Freidel:2017xsi, Freidel:2017wst}, \textit{i.e.} the metastring.
In this description, everything is built out of partonic degrees of freedom 
represented by the entries of the quantum gravitational matrix model, and, 
in the leading term in the expansion involving the fundamental length, 
dark energy is realized as a dynamical geometry of dual spacetime.

The starting point of the metastring formalism is the following worldsheet action \cite{Tseytlin:1990nb,Tseytlin:1990va},
which is chiral, doubles the degrees of freedom (\textit{i.e.} works in phase space), and is manifestly
invariant under Born reciprocity/T-duality:
%and which derives from a chiral world-sheet description (the metastring formulation)
%
\be
S_{2d}\;{=}\;\frac{1}{4\pi}\int_{\Sigma}
\Bigl[\,
  \partial_{\tau}   \X^{A} (\eta_{AB}+\omega_{AB})
- \partial_{\sigma} \X^{A} \,H_{\!AB}
\,\Bigr] \partial_{\sigma} \X^{B}
\;.
\label{e:MSA}
\ee 
Here $\Sigma$ is the worldsheet, the doubled target space variables 
$\X^A = (x^a/\lambda,\tilde{x}_a/\lambda)$ combine 
the sum ($x = x_L+x_R$) and the difference ($\tx = x_L-x_R$) of the left- and right-movers on the string
($a,A=0,1,\cdots,d-1=25$, for the critical bosonic string),
and $\lambda = 1/\epsilon = \sqrt{\alpha'}$ is the string length scale \cite{Polchinski:1998rq}.
The mutually compatible dynamical fields $\omega_{AB}(\X),\eta_{AB}(\X)$, and $H_{\!AB}(\X)$ are respectively:
the antisymmetric symplectic structure $\omega_{AB}$,
the symmetric polarization (doubly orthogonal) metric $\eta_{AB}$, and
the doubled symmetric metric $H_{\!AB}$, which together define a Born geometry \cite{Freidel:2013zga,Freidel:2018tkj,Svoboda:2019fpt}.
See also \cite{Hull:2009mi,Aldazabal:2013sca}.

Quantization renders the doubled ``phase-space'' operators 
$\hat{\X}^A = (\hx^a/\lambda, \htx_b/\lambda)$ inherently non-commutative~\cite{Freidel:2017wst}:
\be
\big[\, \hat{\X}^A,\, \hat{\X}^B\,\big] \,=\, i \omega^{AB}\;.
\label{e:CnCR}
\ee
%
%or, in components, for constant non-zero $\w^{AB}$,
%\bea
%[\hx^a,\htx_b]=2\pi i\l^2 \delta^a_b,\quad
%[\hx^a,\hx^b]=0=[\htx_a,\htx_b],
%\label{e:CnCR1}
%\eea
%where $\l$ denotes the fundamental length scale, such as the Planck scale,
%so that $\e=1/\l$ is the corresponding fundamental energy scale
%Here, the string tension is $\alpha' = \lambda / \epsilon = \lambda^2$.
%This %was 
%fully spacetime covariant formulation was
%found by examining the simplest example of the canonical free string compactified on a circle, in an intrinsically T-duality %covariant formulation of the Polyakov string %.
%and was independently confirmed by examining the algebra of vertex operators in the 2d CFT of a free string compactified %on a circle~\cite{
%Freidel:2013zga, Freidel:2015uug, Freidel:2016pls, 
% Freidel:2017xsi, Freidel:2017wst}.
In this formulation, all effective fields must be regarded a priori as bi-local $\phi(x, \tx)$~\cite{Freidel:2016pls}, subject 
to Eq.~\eqref{e:CnCR}, and therefore inherently non-local (yet covariant) in the conventional $x^a$-spacetime. Such non-commutative field theories \cite{Douglas:2001ba, Grosse:2004yu} generically display a mixing between the ultraviolet (UV) and infrared (IR) physics, 
with continuum limits defined via a double-scale renormalization group (RG) and self-dual fixed points \cite{Grosse:2004yu,Freidel:2017xsi}. 
In the current case, the UV and IR mixing occurs between the observable $x^a$-spacetime and the unobservable
$\tx_a$-spacetime.

The metastring offers a new view on quantum gravity by noting that
the world-sheet can be made modular in our formulation, with the doubling of $\tau$ and $\sigma$,
so that  $\hX (\tau, \sigma)$ can be
in general viewed as an infinite dimensional matrix (the matrix indices coming from the Fourier components of the
doubles of $\tau$ and $\sigma$) \cite{Gopakumar:1994iq,Berglund:2020qcu}.
Then the corresponding metastring matrix model action should look like
\be
S \sim \int  d \tau\, d \sigma\ \mathrm{Tr}
\left[\,
 \partial_{\tau}   \hX^A \partial_{\sigma} \hX^B (\omega_{AB} + \eta_{AB}) 
-\partial_{\sigma} \hX^A \,H_{AB}\, \partial_{\sigma} \hX^B 
\,\right] 
\;,
\ee
where the trace is over the infinite matrix indices. 
The matrix entries become the natural partonic degrees of freedom of quantum spacetime.
The non-perturbative formulation of quantum gravity is
obtained by replacing $\partial_\sigma$ in the above worldsheet action with a
commutator involving one extra $\hX^{26}$ : 
\be
\partial_{\sigma} \hX^A \quad\to\quad \left[\,\hX^{26},\, \hX^A \,\right]\;,\qquad\qquad
A\,=\, 0,\,1,\,\cdots,\,25\;.
\ee
Therefore, as with the relationship between M-theory and type IIA string theory, a fully interactive and non-perturbative formulation of metastring theory is given 
in terms of a matrix model form of the above metastring action (with $a,b,c=0,1,2,\cdots,25, 26$)
\be
S\sim \int d \tau\ \mathrm{Tr} 
\left(\, 
\partial_{\tau} \hX^a \left[ \hX^b, \hX^c \right] \eta_{abc}  
- H_{ac} \left[ \hX^a, \hX^b \right] \left[ \hX^c, \hX^d \right] H_{bd}
\,\right) \;,
\ee
where the first term is of the Chern--Simons form, the second term is of the Yang--Mills form, 
and $\eta_{abc}$ contains both $\omega_{AB}$ and $\eta_{AB}$.
In general, we do not need an overall trace if we think of quantum gravity as a pure quantum theory.
Thus, the following matrix model becomes a pure quantum formulation of quantum gravity
\begin{equation}
\mathbb{S}_{\textit{nc}\text{M}}
\;=\; \frac{1}{4\pi}  
\int_{\tau} %\mathrm{Tr} 
\left(\,
  \partial_{\tau} \hX^{i}
  \left[ \hX^{j}, \hX^{k} \right] g_{ijk} 
- \left[ \hX^{i}, \hX^{j} \right]
  \left[ \hX^{k}, \hX^{\ell} \right] 
  h_{ijk\ell}
\,\right)\;,
\label{metastringAction}
\end{equation}
with $27$ bosonic $\hX$ matrices.\footnote{In this formulation, supersymmetry and its avatars are not fundamental
but emergent \cite{Freidel:2017xsi}.} 
Within this formulation, both matter and gravitational sectors emerge from the
dynamics of the partonic quanta of quantum spacetime.

In particular, in  Ref.~\cite{Berglund:2019ctg} 
it has been argued that the generalized geometric formulation of string theory discussed above,
Eq.~\eqref{metastringAction},
provides an effective description of dark energy, and a de Sitter spacetime.
This is due to the theory's chirality and non-commutatively, as in Eq.~\eqref{e:CnCR}, 
doubled realization of the target space, 
and the stringy effective action on the doubled non-commutative spacetime $(x^a,\tx_a)$,
which leads to the effective action
\be
S_{\text{eff}}^{\textit{nc}}
\;=\;
\int_x \int_{\tilde{x}} \text{Tr} \sqrt{g(x,\tx)}\, 
\Bigl[\,
R(x,\tx) + L_m(x,\tx) + \cdots
\,\Bigr]\;,
\label{e:ncEH}
\ee
where the ellipses denote higher-order curvature terms induced by string theory, and
$L_m$ is the matter Lagrangian put in by hand.
This result can
be understood as a generalization of the famous calculation by Friedan~\cite{rF79b}.  
Owing to Eq.~\eqref{e:CnCR}, we have
\be
\bigl[\,\hx^a, \,\htx_b\,\bigr] \,=\, 2\pi i\,\lambda^2\,\delta^a_b\;,\qquad 
\bigl[\,\hx^a, \,\hx^b \,\bigr] \,=\, 
\bigl[\,\htx_a,\,\htx_b\,\bigr] \,=\, 0\;,
\ee
where $\lambda$ denotes the fundamental length scale, such as the Planck scale,
and $\epsilon = 1/\lambda$ is the corresponding fundamental energy scale, while the string tension is 
$\alpha' = \lambda/\epsilon = \lambda^2$.
Thus $S_{\text{eff}}^{nc}$ expands into numerous terms with different powers of $\lambda$, 
which upon $\tx$-integration, and from the $x$-space vantage point, produce various effective terms.
To lowest (zeroth) order of the expansion in the non-commutative parameter $\lambda$
of $S_{\text{eff}}^{\textit{nc}}$ takes the form:
\bea
S_{d=4} & = & - \int_{x}\int_{\tx} \sqrt{-g(x)} \sqrt{-\tilde{g}(\tx)} 
\,\Bigl[\, 
  R(x) + \tilde{R}(\tx)
\,\Bigr]
\cr
& = & -\int_{x}\sqrt{-g(x)}\left[
R(x)\int_{\tx}\sqrt{-\tilde{g}(\tx)} 
+\int_{\tx}\sqrt{-\tilde{g}(\tx)}\;\tilde{R}(\tx)
\right]
\;,
\label{e:TsSd}
\eea
a result which first was obtained almost three decades ago, effectively neglecting $\omega_{AB}$ 
by assuming that $[\,\hat x^a,\,\htx_b\,] = 0$ \cite{Tseytlin:1990hn}. 
In this leading limit, the $\tx$-integration in the first term of~\eqref{e:TsSd} defines the gravitational constant $G_N$, 
\be
1/G_N \;\sim\; \int_{\tx} \sqrt{-\tilde{g}(\tx)} \;,
\label{OneOverGN}\ee
and in the second term produces a {\it positive} cosmological constant $\Lambda >0$ (dark energy)
\be
\Lambda/G_N \;\sim\; \int_{\tx} \sqrt{-\tilde{g}(\tx)}\,\tilde{R}(\tx) \;.
\label{LambdaOverGN}
\ee
Thus the weakness of gravity is determined by the size of the canonically conjugate dual $\tx$-space, 
while the smallness of the cosmological constant is given by its curvature $\tilde{R}$. 
Ref.~\cite{Berglund:2019ctg} also discusses a see-saw formula for the cosmological constant, as well
as its radiative stability in the underlying general framework of a
non-commutative generalized geometric phase-space formulation of string 
theory~\cite{Freidel:2013zga, Freidel:2015uug, Freidel:2016pls, Freidel:2017xsi, Freidel:2017wst}, 
which is non-local but covariant.

To summarize, a non-perturbative formulation of
quantum gravity can be given in terms of a doubled matrix model, Eq.~\eqref{metastringAction}, 
in which everything is built out of partonic degrees of freedom represented by the entries of doubled matrices $\hX$. 
In the leading term of the effective spacetime description, 
dark energy is realized as a dynamical geometry of the dual spacetime, and consequently, 
is inherently time dependent.

%%%%%%%%%%%%%%%%%%%%%%%%%%%%%%%%%%%%%%%%%%%%%%%%%%%%%%%%%%%%%%%%%%%%%%%%%%%%%% 
\section{Quantum Gravity and Infinite Statistics} 

Matrix models, in the limit of large matrix size, can be directly related to infinite statistics \cite{Gopakumar:1994iq}.
Given that the metastring action, Eq.~\eqref{metastringAction}, formulates
non-perturbative quantum gravity as a matrix model, in which dark energy is realized as the dynamical geometry of 
dual spacetime in the commutative limit,
we argue in this section that infinite statistics can be used to model the fine structure of dark energy \cite{Jejjala:2007hh}. 
In particular, the partonic degrees of freedom of the matrix model, out of which both spacetime and matter degrees of freedom
emerge, obey infinite statistics, and thus, infinite statistics controls the fine structure of dark energy.
%
%In order the implement the above proposal for the origin of dark energy and obtain an explicit formula for a time dependent $\Lambda$, we turn to a general argument based on the role of statistics in quantum gravity. 
The idea here is that by using the general statistical arguments,
we can illustrate the time dependence of $\Lambda$ based on a dynamical dual spacetime geometry
without appealing to any particular models of that dynamics.

%Note that the canonical statement concerning quantum gravitational effects is that they are tied to the Planck scale. Our essential point here is that quantum gravity can be revealed at macroscopic scales via quantum statistics. This observation is a very natural extension of the textbook knowledge regarding the macroscopic realization of quantum statistics, as seen, for example, in the case of the black body radiation.
The proposal that quantum statistical effects are essential in the macroscopic realizations of
quantum gravity has been made in the past. 
First, it was argued in Ref.~\cite{Strominger:1993si} that black hole statistics is infinite statistics \cite{Doplicher:1971wk, Greenberg:1989ty}.
(See also, Ref.~\cite{Minic:1997ym}). 
Also, in Ref.~\cite{Horava:2000tb} a statistical argument was used to argue for probable values of the cosmological constant.
More recently, such statistical arguments were used in Ref.~\cite{Bianchi:2018ula} to 
analyze black hole spin in gravitational wave observations.

Given our proposal regarding the realization of dark energy
in a fundamentally non-local but Lorentz covariant formulation of quantum gravity, on a purely quantum level one should consider the statistics of quanta from which dual spacetime
emerges at large distances.
If one remembers that only one statistics is consistent with non-locality and Lorentz symmetry, both of which
underpin this approach to quantum gravity, one is led to infinite statistics \cite{Greenberg:1989ty, Strominger:1993si} and a fine structure for dark energy.
% The most important point for our current discussion is that this inherently non-local formulation is manifestly covariant, and thus the only known quantum statistics that is consistent with such non-locality (and the CPT theorem and Lorentz symmetry) is infinite statistics \cite{Greenberg:1989ty}. 
Therefore the natural implementation of the physical effects associated
with infinite statistics in the context of dark energy should be sought in this generic non-commutative formulation
of string theory.
%\red{(This paragraph is a bit repetitive.)}

If dark energy originates from the curvature of the dual space, then in the context
of quantum gravity it possesses fine structure.
That fine structure can be deduced from the infinite statistics of 
the quanta of dual spacetime.
The virtue of the metastring action is that supplies a mechanism for UV/IR mixing.
If we simply assume this mixing \textit{ab initio}, our conclusions are generic.

In Ref.~\cite{Jejjala:2007hh} (see also \cite{Jejjala:2007rn}) we have presented a
general argument for the relevance of
infinite statistics for the fine structure of dark energy. 
We remind the reader that infinite statistics is defined in terms of the Cuntz algebra 
\be
\ha_i \ha^{\dagger}_j \;=\; \delta_{ij}\;,
\ee 
which can be viewed as the $q=0$ deformation of the $q$-deformed commutation relations
\be
\ha_i \ha^{\dagger}_j -  q\ha^{\dagger}_i \ha_j \;=\; \delta_{ij}\;.
\ee
The case $q=1$ corresponds to Bose--Einstein statistics, and $q=-1$ to Fermi--Dirac statistics.
Unlike the bosonic and fermionic statistics, infinite statistics realizes any permutation 
(not just even or odd) of the associated $SU(N)$ Young tableaux for $N$ particles.
In particular, infinite statistics governs the master fields of large-$N$ matrix models \cite{Gopakumar:1994iq},
and thus it is appropriate for our approach to quantum gravity based on a non-perturbative 
matrix model formulation. For example, the master field of the quadratic single matrix model is
given as $\ha + \ha^{\dagger}$, with $\ha$ and $\ha^{\dagger}$ satisfying the Cuntz algebra $\ha \ha^{\dagger} =1$ 
\cite{Gopakumar:1994iq}.

More concretely, infinite statistics is quantum Boltzmann statistics, and thus in the quantum 
context, it is of the Wien type \cite{Jejjala:2007hh}. This turns out to be
crucial in our application of infinite statistics to the fine structure of dark energy.
%\red{(Dj: To summarize, in this section we have argued that, given the fact that non-perturbative quantum gravity is formulated as a matrix model, in which, in a certain limit, dark energy is realized as a dynamical geometry of dual spacetime, and given the fact that matrix models, in the limit of the large matrix size, can be directly related to infinite statistics, infinite statistics can be used to model the fine structure of dark energy. Infinite statistics is quantum statistics of distinguishable objects (partons of the quantum gravitational matrix model) and as such it is essentilly just the Boltzmann statistics, or equivalently, the Wien statistics of partonic quanta of spacetime.)}

%%%%%%%%%%%%%%%%%%%%%%%%%%%%%%%%%%%%%%%%%%%%%%%%%%%%%%%%%%%%%%%%%%%%%%%%%%%%%% 
\section{Infinite Statistics and Dark Energy} 

In this section, we use the insight that infinite statistics controls the fine structure of dark energy
in quantum gravity. Infinite statistics is the quantum statistics of distinguishable partons of the quantum gravitational matrix model,
and as such it is essentially just the Boltzmann statistics, or equivalently, the Wien statistics of quantum spacetime partons. 
Given the dual relationship between the observed $x^a$-spacetime and dual $\tx_a$-spacetime, quantum gravity 
is endowed with both UV and IR cut-offs, and thus, the Wien distribution of spacetime quanta/partons responsible for
the fine structure of dark energy comes with an explicit cut-off.

Therefore, motivated by the above general reasoning about the role of infinite statistics in quantum gravity,
let us examine the dark energy spectral function of the quantum Boltzmann (or Wien) type
in the dual energy $\Edual$-space \cite{Jejjala:2007hh}:
\be
\rho_\mathrm{dark\,energy} (\Edual, E_0) \;=\; A\,\Edual^3\, e^{-B \Edual/E_0}\;,
\ee
where $A$ and $B$ are dimensionless constants.
From Eq.~\eqref{LambdaOverGN}, we have
\be
\rho_{vac}(\Edual_{UV}) \;=\; \dfrac{\Lambda(\Edual_{UV})}{8\pi G_N}
\;=\; \int_0^{\Edual_{UV}} d\Edual\ \rho_\mathrm{dark\,energy}(\Edual, E_0)
\;,
\ee
where $\rho_{vac}(\Edual_{UV})$ is the effective vacuum energy density in the observable spacetime,
while $\Edual_{UV}$ is the UV cutoff in the unobservable dual spacetime.
Due to the UV/IR correspondence between the two spacetimes, we have
\be
\Edual_{UV}E_{IR} \;=\; \mu\;,
\ee
where $E_{IR}$ is the IR cutoff in the observable spacetime, and $\mu$ is an invariant associated
with the doubly orthogonal group of transformations in the metastring approach \cite{Freidel:2013zga, Freidel:2015uug, Freidel:2016pls, Freidel:2017xsi, Freidel:2017wst}.
We expect $E_{IR}$ to be governed by the size of the observable Universe, thus
\be
E_{IR} \;=\; \dfrac{E_0}{a} \;=\; E_0(1+z)\;,
\ee
where $a$ is the scale factor of the Friedmann--Lema\^{i}tre--Robertson--Walker metric,
$z$ is the redshift, and we identify $E_0$ as the current ($z=0$) IR cutoff.
Thus,
\be
\Edual_{UV} \;=\; \dfrac{\mu}{E_{IR}} \;=\; \dfrac{\mu a}{E_0} \;=\; \dfrac{\mu}{E_0(1+z)}\;,
\ee
and we find
\be
\rho_{vac}(z) 
\;=\; \dfrac{\Lambda(z)}{8\pi G_N}
\;=\; \int_{0}^{\Edual_{UV}}d\Edual\ \rho_\mathrm{dark\,energy}(\Edual, E_0)
\;=\; \rho_*\Bigl[ 1 - b(\xi) \Bigr]
\;,
\ee
where
\be
\rho_*\,=\, \dfrac{6A}{B^4} E_0^4\;,\qquad
b(\xi) \,=\, \left(1 + \xi + \dfrac{\xi^2}{2} + \dfrac{\xi^3}{6}\right) e^{-\xi}\;,
\ee
and
\be
\xi 
\,=\, \dfrac{B\Edual_{UV}}{E_0}
\,=\, \dfrac{B\mu}{E_0^2(1+z)}
\,=\, \dfrac{\xi_0}{1+z}\;,\qquad
\xi_0 \,=\, \dfrac{B\mu}{E_0^2}\;.
\ee
The proportionality of $\rho_*$ to $E_0^4$ is analogous to the derivation of the Stefan--Boltzmann $T^4$ law
from the Wien distribution \cite{Jejjala:2007hh}.
The functional form of $b(\xi)$ is shown in Figure~1(a).
Therefore, in our proposal, $\Lambda(z)$ evolves as
\be
\dfrac{\Lambda(z)}{\Lambda(0)}
\,=\, \dfrac{1-b(\xi)}{1-b(\xi_0)}\;.
\label{Lambda-z}
\ee
In the limit $z\to\infty$ (early Universe), we have $\xi = \xi_0/(1+z)\to 0$, 
$b(\xi)\to 1$, and $\Lambda(z)/\Lambda(0) \to 0$.
The $z$ dependence of this ratio in the range $z\le 1$ 
is shown for several values of $\xi_0$ in Figure.~1(b).

\begin{figure}[t]
\begin{center}
\subfigure[]{\includegraphics[width=7cm]{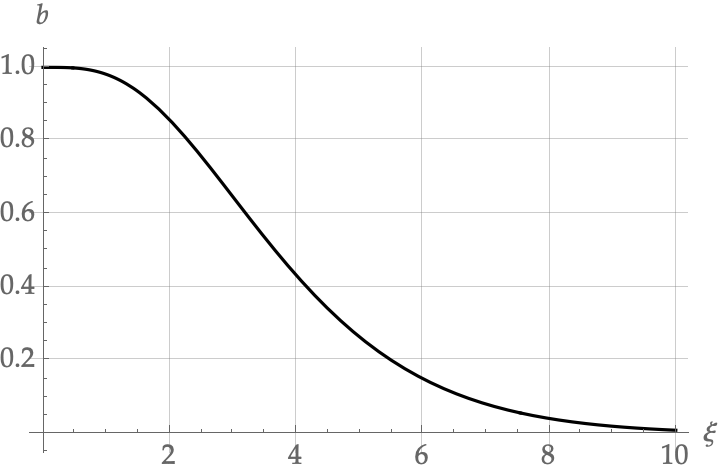}}
\hspace{1cm}
\subfigure[]{\includegraphics[width=7cm]{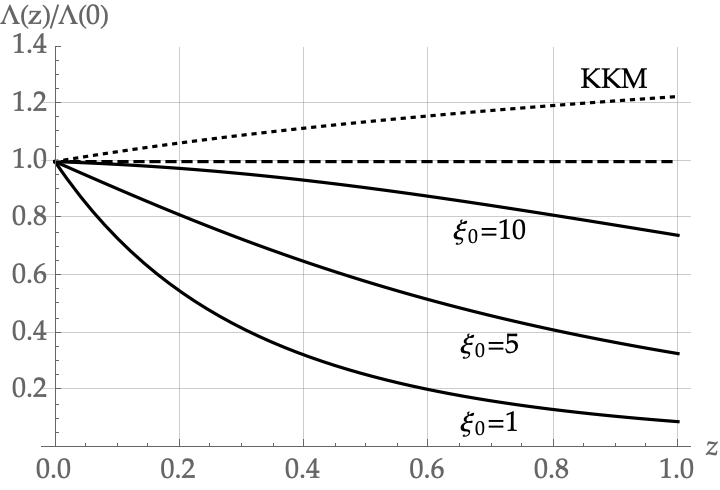}}
\caption{\textsf{(a) Behavior of the function $b(\xi)$.
(b) The behavior of $\Lambda(z)/\Lambda(0)$ in our proposal for $\xi_0=1$,
$\xi_0=5$, and $\xi_0=10$ compared against constant $\Lambda$ (dashed) and that 
from Ref.~\cite{Kitamoto:2019rij} (dotted).}}
\label{b-graph}
\end{center}
\end{figure}

What can the value of the invariant $\mu$ be?
We can identify $E_{IR}=E_0$ with the current vacuum energy scale.
If we set the curent $\Edual_{UV}$ to the Planck mass $E_P = G_N^{-1/2}$, c.f.
Eq.~\eqref{OneOverGN}, we have 
\bea
\mu 
& = & E_{IR}\Edual_{UV}
\;=\; E_0 E_P
\vphantom{\Big|}\cr
& = & \left(2.24\times 10^{-12}\ \text{GeV}\right)\left(1.22\times 10^{19}\ \text{GeV}\right) 
\;=\; 2.73\times 10^7\ \text{GeV}^2 
\;.
\eea
This means $\sqrt\mu \simeq 5.23\ \text{TeV}$, more or less the scale at which the LHC operates.
However, this is only a very special choice, and in general, $\mu$ is a parameter we should fit.

%%%%%%%%%%%%%%%%%%%%%%%%%%%%%%%%%%%%%%%%%%%%%%%%%%%%%%%%%%%%%%%%%%%%%%%%%%%%%% 
\section{Evolution of the Hubble Parameter}

\begin{figure}[ht]
\begin{center}
\includegraphics[width=7cm]{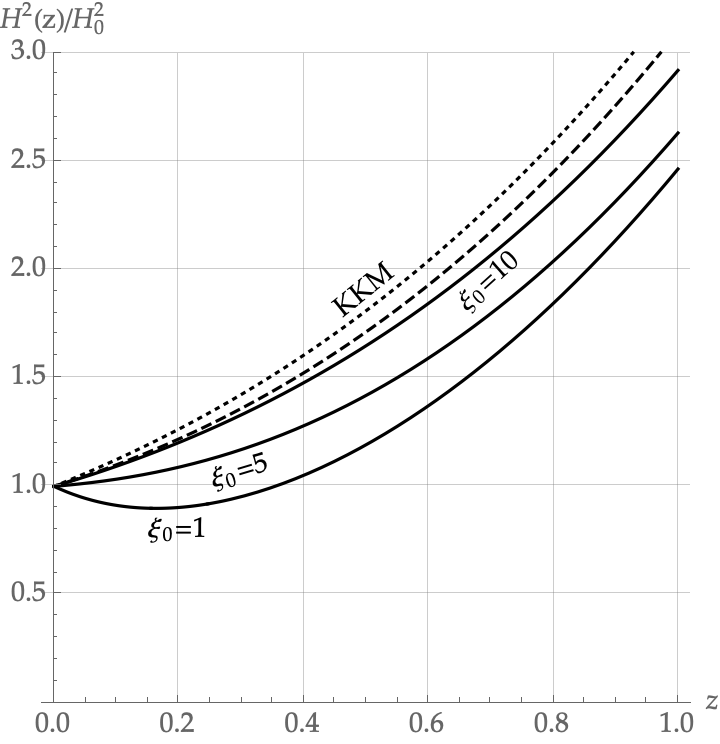}
\caption{\textsf{The behavior of $H^2(z)/H_0^2$ in our proposal for $\xi_0=1$,
$\xi_0=5$, and $\xi_0=10$ compared against the constant $\Lambda$ case (dashed) and that 
from Ref.~\cite{Kitamoto:2019rij} (dotted).}}
\end{center}
\label{H2-graph}
\end{figure}

Let us see how the $z$ dependence of $\Lambda(z)$ will affect the evolution of the Hubble parameter $H(z)$.
$H(z)$ evolves in a spatially flat ($k=0$) matter dominated ($z\alt 3000$) universe as \cite{Weinberg:2008zzc}
\begin{equation}
H^2(z) \;\equiv\; \left(\dfrac{\dot{a}}{a}\right)^2 \,=\, H_0^2\Big[
\Omega_m(1+z)^3 + \Omega_\Lambda(z)
\Bigr]\;,
\label{HubbleParameterEvolution}
\end{equation}
where $\Omega_\Lambda(z)$ and $\Omega_m$ are respectively the 
density fractions of vacuum energy and matter normalized to the present critical density $\rho_{c,0} = 3H_0^2/8\pi G_N$:
\begin{equation}
\Omega_\Lambda(z) \;=\; \dfrac{\rho_\Lambda(z)}{\rho_{c,0}} \;=\; \dfrac{\Lambda(z)}{3H_0^2}\;,\qquad
\Omega_m\;=\; \dfrac{\rho_m}{\rho_{c,0}}\;.
\end{equation}
The Hubble constant $H_0$ is the value of the Hubble parameter $H(z)$ at redshift $z=0$.
Consistency of Eq.~\eqref{HubbleParameterEvolution} requires
\begin{equation}
\Omega_m + \Omega_\Lambda(0) \;=\; 1\;.
\end{equation}
Observations yield $\Omega_M = 0.3$ and $\Omega_{\Lambda}(0) = 0.7$ \cite{Aghanim:2018eyx}.
Therefore,
\be
\frac{H (z)^2}{H_0^2} \;=\; 
\Omega_{M} (1+z)^3 
\,+\,
\Omega_{\Lambda}(0)\left[\dfrac{\Lambda(z)}{\Lambda(0)}\right]
\;.
\ee
Substituting Eq.~\eqref{Lambda-z} into this expression will give us the $z$ dependence of $H(z)$.
The behavior of $H^2(z)/H_0^2$ is shown for several values of $\xi_0$ in Figure~2, compared against
the constant $\Lambda$ case.  
Note that in our proposal, the $\Lambda(z)$ contribution to $H(z)$ vanishes when $z\gg 1$.
Also shown in Figure~2 is the result of Kitamoto, Kitazawa, and Matsubara (KKM) in Ref.~\cite{Kitamoto:2019rij}
in which the authors compute a $\beta$-function for $g = G_N H^2$ in Einstein gravity in four dimensional de Sitter space to obtain (their formula (5.39))
\be
\left[\frac{H (z)^2}{H_0^2}\right]_\mathrm{KKM} 
\;=\; \Omega_m \,(1+z)^3 + \Omega_\Lambda(0) \log\Bigl[e + \log{(1+z )}\Bigr]\;.
\label{KKM}
\ee
While the KKM model makes certain assumptions about a conformally coupled scalar field to ensure the running of the coupling $g$ and identifies the behavior of the Hubble constant with quantum IR effects, we motivate the time evolution of $\Lambda(z)$ from infinite statistics and the dynamical mixing of UV and IR degrees of freedom.
The evolution of the Hubble parameter in the context of the model proposed here differs significantly from either the constant $\Lambda$ case or the KKM model for most values of $\xi_0$. Thus there exists the potential for a definitive prediction to be made of the evolution Hubble parameter by fitting the model proposed here to current cosmological data. We will discuss the phenomenology of the above formul\ae\ for $H^2(z)/H_0^2$,
and whether they have any relevance to the $H_0$ tension elsewhere \cite{jkmt}.

%%%%%%%%%%%%%%%%%%%%%%%%%%%%%%%%%%%%%%%%%%%%%%%%%%%%%%%%%%%%%%%%%%%%%%%%%%%%%% 
%\paragraph{Concluding remarks:} 
\section{Concluding Remarks}

In this paper we have discussed the relation between infinite statistics and dynamical dark energy  
based on the recent proposal for the origin of dark energy from the curvature of dual spacetime \cite{Berglund:2019ctg}
in the context of the new approach to quantum gravity of 
Refs.~\cite{Freidel:2013zga, Freidel:2015uug, Freidel:2016pls, Freidel:2017xsi, Freidel:2017wst}.
Specifically, following the general arguments for the relevance of infinite statistics in quantum
gravity \cite{Jejjala:2007hh}, we have derived a formula for $\Lambda(z)$ in a particular example 
based on the quantum statistical effects (due to infinite statistics) within this general approach.

Note that we have not included the matter sector explicitly in the above discussion. 
However, the dual part of the matter sector can be naturally related
to the dark matter sector that is sensitive to dark energy \cite{Ho:2010ca}
which illustrates the unity of the description of the entire dark sector based on the properties of
the dual spacetime, as predicted by the above generic non-commutative formulation of 
string theory/quantum gravity~\cite{Freidel:2013zga, Freidel:2015uug, Freidel:2016pls, Freidel:2017xsi, Freidel:2017wst}.

%%%%%%%%%%%%%%%%%%%%%%%%%%%%%%%%%%%%%%%%%%%%%%%%%%%%%%%%%%%%%%%%%%%%%%%%%%%%%% 
\section*{Acknowledgments}
We thank P.~Berglund, E.~Bianchi,  L.~Freidel, S.~Horiuchi, T.~Hubsch and R.~G.~Leigh  
for helpful discussions. 
VJ is supported by the Simons Foundation Mathematical and Physical Sciences Targeted Grants to Institutes, Award ID:509116 and the South African Research Chairs Initiative of the Department of Science and Technology and the National Research Foundation.  
DM and TT are supported in part by the US Department of Energy (DE-SC0020262, Task C). 
DM is also supported by the Julian Schwinger Foundation.
TT is also supported in part by the US National Science Foundation (PHY-1413031).

%%%%%%%%%%%%%%%%%%%%%%%%%%%%%%%%%%%%%%%%%%%%%%%%%%%%%%%%%%%%%%%%%%%%%%%%%%%%%% 
%: Bibliography
\footnotesize\baselineskip=2.5ex
%
%\bibliographystyle{utphys}
%\bibliography{RefsHR}
%\end{document}
%
%\clearpage

\end{document}